\documentclass[prb,twocolumn,amsmath,amssymb,showpacs,superscriptaddress]{revtex4}

\usepackage{graphicx}
\usepackage{dcolumn}
\usepackage{bm}

\begin{document}


\title{Long-time electron spin storage via dynamical suppression of\\
hyperfine-induced decoherence in a quantum dot}

\author{Wenxian Zhang}

\affiliation{Ames Laboratory, Iowa State University, Ames, IA
50011, USA}
\affiliation{Department of Physics and Astronomy,
Dartmouth College, Hanover, NH 03755, USA}

\author{N. P. Konstantinidis}
\altaffiliation{Present address: Institut f\"{u}r Festk\"{o}rper-Theorie
III, Forschungszentrum J\"{u}lich, 52425 J\"{u}lich, Germany, and Institut
f\"{u}r Theoretische Physik A, Physikzentrum, 52056 Aachen, Germany}

\affiliation{Ames Laboratory, Iowa State University, Ames, IA 50011,
USA}

\author{V. V. Dobrovitski}
\affiliation{Ames Laboratory, Iowa State University, Ames, IA 50011,
USA}

\author{B. N. Harmon}
\affiliation{Ames Laboratory, Iowa State University, Ames, IA
50011, USA}

\author{Lea F. Santos}

\affiliation{Department of Physics, Yeshiva University, New York, NY 10016, USA}

\author{Lorenza Viola}
\affiliation{Department of Physics and Astronomy, Dartmouth
College, Hanover, NH 03755, USA}

\date{\today}

\begin{abstract}
The coherence time of an electron spin decohered by the nuclear spin
environment in a quantum dot can be substantially increased by
subjecting the electron to suitable dynamical decoupling sequences.
We analyze the performance of high-level decoupling protocols by using
a combination of analytical and exact numerical methods, and by paying
special attention to the regimes of large inter-pulse delays and
long-time dynamics, which are outside the reach of standard average
Hamiltonian theory descriptions.  We demonstrate that dynamical
decoupling can remain efficient far beyond its formal domain of
applicability, and find that a protocol exploiting concatenated design
provides best performance for this system in the relevant parameter
range.  In situations where the initial electron state is known,
protocols able to completely freeze decoherence at long times are
constructed and characterized.  The impact of system and control
non-idealities is also assessed, including the effect of intra-bath
dipolar interaction, magnetic field bias and bath polarization, as
well as systematic pulse imperfections. While small bias field and small
bath polarization degrade the decoupling fidelity, enhanced performance
and temporal modulation result from strong applied fields and high
polarizations. Overall, we find that if the relative errors of the
control parameters do not exceed 5$\%$, decoupling protocols can still
prolong the coherence time by up to two orders of magnitude.
\end{abstract}

\pacs {03.67.Pp, 76.60.Lz, 03.65.Yz, 03.67.Lx}

\maketitle

\section{Introduction}

Electron and nuclear spin degrees of freedom are promising candidates
for a variety of quantum information processing (QIP) devices
\cite{DiVincenzo95,AwschalomSciAm02,Baugh07}.  While the wide range of
existing microfabrication techniques make solid-state architectures
extremely appealing in terms of large-scale integration, such an
advantage is seriously hampered by the noisy environments which are
typical of solid-state systems, and are responsible for unwanted rapid
decoherence.  For an electron spin localized in a GaAs quantum dot
(QD) \cite{Loss98}, for instance, the relevant coherence time is
extremely short: at typical experimental temperatures (T $\sim
100$~mK) and sub-Tesla magnetic fields, the electron free induction
decay (FID) time $T_2^*\sim 10$~ns \cite{Petta05,Johnson05,Koppens05},
the dominant decoherence mechanism being the hyperfine coupling to the
surrounding bath of Ga and As nuclear spins.  Although
efforts for achieving faster gating times may contribute to alleviate
the problem, it remains both highly desirable and presently more
practical to extend the coherence time of the central system (the
electron spin) in the presence of the spin bath.

Several proposals have been recently put forward to meet this
challenge. A first strategy is to manipulate the spin bath.
Polarizing the nuclear spins, for instance, may significantly
increase the coherence time \cite{Burkard99,Deng06}, provided that
nuclear-spin polarization $\gtrsim 99\%$ may be achieved.  This,
however, remains well beyond the current experimental
capabilities.  Another suggestive possibility, based on narrowing
the distribution of nuclear spin
states~\cite{Coish04,Stepanenko06,Klauser06}, has been predicted
to enhance electron coherence by up to a factor of hundred, upon
repeatedly measuring and pumping the electron into an auxiliary
excited trion state -- which also appears very challenging at
present.

As an alternative approach, direct manipulation of the central spin by
means of electron spin resonance (ESR) \cite{Gerstein85} and dynamical
decoupling (DD) \cite{Haeberlen76,Slichter92,Viola98,Viola99} techniques appears ideally suited to hyperfine-induced decoherence suppression,
in view of the long correlation time and non-Markovian behavior which distinguish the nuclear spin reservoir. A single-pulse Hahn-echo protocol
has been implemented recently in a double-QD device \cite{Petta05},
increasing the coherence time by two orders of magnitude. Significant potential of more elaborated pulse sequences, such as the multi-pulse
Carr-Purcell-Meiboom-Gill (CPMG) protocol \cite{Witzel07} and
concatenated DD (CDD)
\cite{Khodjasteh05,Khodjasteh06,Bergli06,Yao07}, has been established
theoretically for a single QD subjected to a {\em strong} external
bias field, whereby the electron effectively undergoes a purely
dephasing process.  The DD problem for the more complex situation of a
{\em zero or low} bias fields, where pure dephasing and relaxation
compete, has been recently examined in Refs.
\onlinecite{Zhang07a, Zhang07b}.
Having established the existence of highly effective DD schemes for
electron spin storage, the purpose of this work is twofold:
first, to gain a deeper understanding of the factors influencing DD
performance and the range of applicability of conclusions based on
analytical average Hamiltonian theory (AHT) approaches; and second,
to assess the influence of various factors which may cause the
system and/or control Hamiltonians to differ from the idealized
starting point chosen for analysis.

Aside from its prospective practical significance, developing and
benchmarking strategies for decoherence suppression in various spin
nanosystems is interesting from the broader perspective of quantum
control theory.  In particular, standard theoretical tools usually
employed for the analysis of DD performance, such as AHT and the
Magnus Expansion (ME), have very restrictive formal requirements of
applicability (very fast control time scales, bounded environments,
etc), which may be hard to meet in realistic systems.  Thus, in-depth
studies of physically motivated examples are essential to understand
how to go beyond the formal error bounds {\em sufficient} for convergence, and to identify more realistic {\em necessary}
criteria for DD efficiency.
In this sense, a QD system, described by the central spin model
(a central spin-$1/2$ interacting with a bath of $N$ external spins
\cite{Gaudin76,Prokofev00}) both provides a natural testbed for
detailed DD analysis, and paves the way to understanding more complex
many-spin central systems.

In this work, we present a quantitative investigation of DD as a
strategy for robust long-time electron spin storage in a QD. The
content of the paper is organized as follows. In
Sec.~\ref{sec:formulation}, we lay out the relevant control setting,
by describing the underlying QD model as well as the deterministic and
randomized DD protocols under consideration. Among periodic schemes,
special emphasis is devoted to the concatenated protocol (PCDD$_2$),
which was identified as the best performer for this system
in Ref.~\onlinecite{Zhang07a}.  Exact AHT results are obtained up to
second-order corrections in the cycle time, which allow additional
physical insight on the underlying averaging and on DD-induced {\em
bath renormalization} to be gained. Details
on the methodologies followed to assess the quality of DD and to
effect exact numerical simulations of the central spin coupled to up
to $N=25$ bath spins are also included in Sec.~\ref{sec:formulation}.

In Sec.~\ref{sec:performance}, numerical results on best- and
worst-case performance of DD protocols for short evolution times
are presented and compared with analytical predictions from AHT/ME
under convergence conditions -- in particular, $\omega_c \tau \ll 1$,
where $\omega_c$ and $\tau$ are the upper cutoff frequency of the
total system-plus-bath spectrum, and the time interval between (nearly instantaneous) consecutive control operations, respectively.
Evolution times as long as $\sim
1000 T_2^*$ for $\tau \lesssim T_2^*$ are able to be investigated
numerically, other decoherence mechanisms becoming relevant for yet
longer times. In the best-case scenario, where decoherence of a known
initial state may be frozen under appropriate cyclic DD protocols
\cite{Zhang07a}, the dependence of the attainable asymptotic coherence
value on $\tau$ is elucidated in the small $\tau$ region.
Sec.~\ref{sec:real} is devoted to further investigating the effect of
experimentally relevant system features and/or non-idealities, such as
the presence of residual dipolar couplings between the nuclear spins,
the influence of an applied bias magnetic field, and the role of
initial bath polarization. In Sec.~\ref{sec:ci}, the effect of
systematic control imperfections such as finite width of pulses and
rotation angle errors is quantitatively assessed. We present
conclusions in Sec.~\ref{sec:conclusion}.


\section{System and control assumptions}
\label{sec:formulation}

In this section we describe the model spin-Hamiltonian for a
typical semiconductor QD, and typical picture of the decoherence
dynamics of an electron spin in a QD. We present the DD methods
used to suppress the electron spin decoherence and the metrics for
DD performance, followed by numerical methods we employed.

\subsection{QD model Hamiltonian}

The decoherence dynamics of an electron spin $\bf S$ localized in a QD
and coupled to a mesoscopic bath consisting of $N$ nuclear spins ${\bf
I}_k$, $k=1,\dots N$, may be accurately described by the effective
spin Hamiltonian
\cite{Dobrovitski03,Al-Hassanieh06,Zhang06,Taylor06,Zhang07a,Zhang07b}
\begin{equation}H = H_S + H_B + H_{SB}, \label{eq:h}
\end{equation}
where, in units $\hbar=1$,
\begin{eqnarray}
H_S &=& \omega_0 S_z, \label{Hel} \\
H_B &=& \sum_{k=1}^N \sum_{\ell < k =1}^N \Gamma_{k\ell} ({\bf I}_k
\cdot {\bf I}_\ell-3I^z_k I^z_\ell), \label{Hb} \\
H_{SB} &=& \sum_{k=1}^N A_k {\bf S} \cdot {\bf I}_k, \label{fermi}
\end{eqnarray}
Here, the Hamiltonian $H_S$ describes the electron spin,
$\omega_0=g_e^*\mu_BB_0$ being the Zeeman splitting in an external
magnetic field $B_0$, $g_e^*$ is the effective Land\'e factor of
the electron spin, and $\mu_B$ is the Bohr's magneton. The
Hamiltonian $H_B$ is the bath Hamiltonian, describing dipolar
interactions with strength $\Gamma_{k\ell}$ between nuclear spin
$k$ and $\ell$.  For our analysis, a nearest-neighbor
approximation to $H_B$ is adequate, with $\Gamma_{k\ell}$ taken as
random numbers uniformly distributed between
$[-\Gamma_0,\Gamma_0]$, so that the parameter $\Gamma_0$
upper-bounds the strength of intra-bath couplings. The relevant
system-bath coupling Hamiltonian $H_{SB}$ accounts for the Fermi
contact hyperfine interaction, the coupling parameter
$A_k=(8\pi/3)g_e^*\mu_Bg_n\mu_n u^2({\mathbf x}_k)$ being
determined by the electron density $u^2({\mathbf x}_k)$ at the
$k$-th nuclear spin site ${\mathbf x}_k$ and by the Land\'e factor
of the nuclei $g_n$. Other small contributions to the total QD
Hamiltonian, such as the Zeeman splitting of the nuclear spins and
anisotropic electron-nuclear couplings, may be neglected for the
current purposes~\cite{Zhang06,Taylor06} (see, however,
Ref.~\onlinecite{Hodges07} for recent developments on
controllability in the presence of anisotropic hyperfine
couplings).

Upon tracing over the nuclear-spin reservoir, the electron spin
described by Eq.~(\ref{eq:h}) undergoes fast decoherence with a
characteristic FID time of
$$T_2^* \simeq \frac{1}{A}
\sqrt{\frac{8}{N} \frac{3}{4 I(I+1) } }, \;\;\;\;
A=\sqrt{\frac{\sum_k A_k^2}{N}},  $$
\noindent
where $A \approx 10^{-4}\mu$eV for typical GaAs QDs with
$N=10^6$ $I_k=I=3/2$ nuclear spins~\cite{Paget77,Zhang06}. This
results in a $T_2^*$ value of about $10$ ns
\cite{Johnson05,Koppens05,Petta05,Taylor06}, which is too short
for QIP applications.  In simulations, we shall neglect the
$I$-dependence of the FID time and simply set the nuclear spin
value to $I_k=1/2$ for all $k$. The FID time also turns out to
depend very weakly on the applied bias field $B_0$ as long as
$B_0$ is smaller than or comparable to the Overhauser field of the
unpolarized nuclear spins~\cite{Zhang06}.  A similar conclusion
holds for a weakly polarized nuclear spin bath, with polarization
$p\ll 1$. Throughout this paper, energies shall be expressed in
units of $A$, and time shall be expressed in units of $1/A$. Since
$\Gamma_0\ll A_k$ for typical GaAs QDs~\cite{Taylor06}, we shall
take $\Gamma_0=0$ (hence $H_B=0$) unless explicitly stated.

\subsection{QD decoupling protocols}

Compared to other decoherence control strategies, DD has many attractive
features: it is a purely open-loop control method which, as such,
avoids the need of measurements and/or feedback; it does not rely on a
particular initial state which might be hard to prepare; its design and
implementation may significantly benefit from the extensive expertise
available from nuclear magnetic resonance (NMR) and ESR techniques.

We shall first assume to have access to {\em ideal} control resources,
and defer the discussion of control limitations to Sec.~\ref{sec:ci}.
In this idealized setting, DD of the electron spin is implemented by
subjecting it to a sequence of {\em bang-bang}
$\pi_{\hat{\bm{n}}}$ pulses \cite{Viola98}, each
instantaneously rotating the spin ${\mathbf S}$ along a given control
axis $\hat{\bm{n}}$ by an angle $\pi$. Consecutive pulses are
separated by a time interval $\tau$, resulting in a total
time-dependent Hamiltonian of the form
\begin{eqnarray}
H_{tot}(t) &=& H_c(t) + H , \nonumber \\
H_c(t) & =&  \sum_\ell \pi ({\bf S}\cdot {\hat{\bm{n}}_\ell})
\delta(t-\ell\tau),
\end{eqnarray}
where $\delta(\cdot)$ is the delta function, $\ell=1,2,\dots$ labels
the applied pulses, and $H$ is given by Eq.~(\ref{eq:h}). The evolution
of the coupled electron-nuclear system in the physical frame is then
described by the unitary propagator
\begin{eqnarray}
U(t) &=& {\mathcal T}\exp\left[-i\int_0^t H_{tot}(s) ds\right],
\end{eqnarray}
where $\mathcal T$ denotes as usual temporal ordering.

\subsubsection{Deterministic single-axis DD}

DD protocols involving control rotations of the central spin about a
fixed axis achieve {\em selective} averaging of spin components
perpendicular to the rotation axis.  Such sequences are effective when
a preferred direction is present either in the underlying physical
Hamiltonian or in the initial quantum state to preserve. For instance,
in the presence of a strong static bias field, $B_0 \gtrsim 1$ T,
electron spin flips are energetically suppressed, and pure dephasing
in the transverse direction is the dominant decoherence
source~\cite{Shenvi05,deSousa05,Khodjasteh06,Witzel07}.  In such a
situation, the interaction Hamiltonian Eq.~(\ref{fermi}) simplifies to
\begin{equation}
H_{\rm eff} = S_z \sum_k A'_k I_k^z ,
\end{equation}
with renormalized coupling constants $A'_k$. The $z$-component of the
central spin remains constant, while the transverse component $s_\perp
= \sqrt{\langle S_x\rangle^2 + \langle S_y\rangle^2}$ undergoes
Gaussian FID with time constant $T_2^*$. By applying a $\pi_x$ pulse
at times $t=\tau$ and $2\tau$ (the so-called Hahn echo, for brevity
denoted as [$\tau X\tau X$]), the electron spin is refocused
completely~\cite{Slichter92} at time $T=2\tau$,
\begin{eqnarray}
U(2\tau) &=& e^{-i\pi S_x} e^{-iH_{\rm eff}\tau} e^{-i\pi S_x}
e^{-iH_{\rm eff}\tau} =
-{\mathbf 1},  \nonumber \\
S_{x,y}(2\tau) &=& U^\dag (2\tau)S_{x,y} U(2\tau) = S_{x,y}(0).
\end{eqnarray}
A protocol consisting of repeated spin-echoes shall be referred to as
CPMG henceforth. In the case of zero or small bias field where the
full Fermi contact interaction is relevant (see Sec.~\ref{sec:bp} for
more discussion on the case of a nonzero bias field), selective DD no
longer removes the effect of the hyperfine coupling on a {\em generic}
initial electron state.  Rather, to lowest order in $\tau$, the control
has the effect of symmetrizing the system-bath Hamiltonian according
to the direction of the applied pulses~\cite{Viola99,Zanardi99}.  The
presence of such a control-induced approximate symmetrization is essential to understand the possibility of decoherence
freezing~\cite{Zhang07a}, to be further discussed in
Sec.~\ref{sec:fb} as well as in the Appendix.

\subsubsection{Deterministic two-axis DD}
\label{twoaxis}

In the absence of a bias field ($B_0=0$), relaxation in the
longitudinal direction is as important as dephasing. Thus, removing the
effect of the nuclear reservoir on the electron spin is only possible
by using a control protocol which achieves {\em non-selective} (or
universal) DD.  Several sequences have been constructed for
finite-dimensional systems by assuming control over a basic set of
unitary operations forming a discrete group ${\cal G}=\{g_j\}$,
$j=0,1, \ldots, |{\cal G}|-1$ (DD
group)~\cite{Viola98,Viola99,Khodjasteh05}.  In the simplest case of
{\em cyclic DD}, the control propagator is sequentially steered
through the DD group in a predetermined order, the change from $g_i$
to $g_j$ being effected through the application of a bang-bang pulse
$P_{i,j}=g_{j}g_{i}^{\dagger}$ -- for instance, ${\cal G}=\{I,X \}$
in the above-mentioned CPMG protocol, $I$ denoting the identity
operation. Thanks to the inherent periodicity of the control action,
with cycle time $T_c = |{\cal G}|\tau$, the DD analysis can be
naturally carried out within the AHT by invoking the ME
\cite{Haeberlen76,Gerstein85,Slichter92},
\begin{equation}
U(T_c)=\exp(-i \overline {\cal H} T_c), \;\;\; \overline {\cal H}
= \sum_{k=0}^\infty \overline {\cal H}^{(k)}, \label{ME}
\end{equation}
where $\overline {\cal H}$ denotes the AHT.  In principle, arbitrary
high-order contributions $\overline {\cal H}^{(k)}$ may be explicitly
evaluated by knowing the applied control sequence \cite{tree}.
A DD scheme for an open-system
Hamiltonian of the form (\ref{eq:h}) is said to be of order $m$
if $\overline {\cal H}^{(0)}= H_B$ and the first non-zero contribution
to $\overline {\cal H}$ arises from $\overline {\cal H}^{(m)}$,
thereby being of order ${\mathcal O}(T_c^m)$ in the expansion parameter
$T_c$.  Estimates of the convergence radius for the ME depend
sensitively on how the strength of $H$ is quantified, which may be
especially delicate for mesoscopic- and infinite-dimensional
environments \cite{KLV,Oteo01,Khodjasteh06}. A conservative
convergence bound arises by assuming that a finite upper spectral
cutoff may be identified, $||H|| \sim \omega_c < \infty$, and the
condition $\omega_c T_c < 1$ is
obeyed~\cite{Maricq,Oteo01,Viola05,Khodjasteh06}. A more precise
characterization of sufficient convergence conditions for the ME
has been recently established in Ref.~\onlinecite{Casas}.

{\em Periodic DD} (PDD) is the simplest non-selective cyclic
protocol, which ensures that the unwanted evolution is removed to
first order in the ME at every $T_n=nT_c$, $n \in {\mathbb N}$, for
sufficiently short $T_c$. For a single central spin, PDD is based on
the irreducible Pauli group ${\cal G}_P=\{I,X,Y,Z \}$, where
$X,Y,Z$ are Pauli matrices, up to irrelevant phase
factors~\cite{Viola98,Viola99}.  A possible implementation,
corresponding to the group path $(I,X,Z,Y)$, involves two-axis
control sequences of the form
\begin{eqnarray}
C_1 = \mbox{PDD} &=& e^{-i\pi S_z}C_0e^{-i\pi S_x}C_0e^{-i\pi S_z}C_0
e^{-i\pi S_x}C_0 \nonumber \\
&=& C_0 X C_0 Z C_0 X C_0 Z \label{eq:pdd}
\end{eqnarray}
with $C_0=\exp(-iH\tau)$ representing the operator of free evolution
between pulses. Note that in the second line of Eq.~(\ref{eq:pdd}),
the time convention used for pulse sequence descriptions is followed,
that is, time evolves from left to right.  For such a sequence, the
lowest-order ME reads
\begin{eqnarray}
C_1 &=& \exp\left[-i T_c (\overline{\cal H}^{(0)} +
\overline{\cal H}^{(1)} + \cdots)\right]
\end{eqnarray}
with
\begin{eqnarray}
\overline {\cal H}^{(0)} &\hspace*{-1mm}=\hspace*{-1mm}
& \frac{1}{T_c}\left[H_4 + H_3 + H_2 +
H_1\right ] \tau = H_B, \\
\overline {\cal H}^{(1)} &\hspace*{-1mm}=
\hspace*{-1mm}& -{\frac{i}{2T_c}}\sum_{j=1}^4
\sum_{i=1}^{j} \left[H_j, H_i\right] \tau^2 \nonumber \\
&\hspace*{-1mm}=\hspace*{-1mm}& -S_y {\frac{\tau}{4}}
\sum_k \sum_\ell A_kA_\ell
(I_k^zI_\ell^x+I_k^xI_\ell^z), \label{eq:pddh1}
\end{eqnarray}
where $H_1 = H$, $H_2 = XHX$, $H_3 = YHY$, and $H_4 = ZHZ$,
respectively. Thus, PDD achieves first-order DD, with the unwanted
hyperfine coupling vanishing in the limit $\tau\to 0$, and
$\overline {\cal H}^{(0)}=0$ for a non-dynamical bath. Note that the
decoupling happens after completion of each cycle, so everywhere in
the subsequent analysis of cyclic DD, we shall assume that the
evolution is sampled {\em stroboscopically} at instants $t_j=j T_c,
j=1,2,\dots$

A simple strategy to improve over PDD averaging is based on exploiting
time-symmetrization \cite{Haeberlen76,Viola99} -- leading to so-called
{\em symmetric DD} (SDD).  Corresponding to the above-defined PDD
protocol, the SDD protocol relevant to our problem is defined by the
control cycle
\begin{equation}
\mbox{SDD}= C_0XC_0ZC_0XC_0C_0XC_0ZC_0XC_0\,.
\end{equation}
Symmetrization guarantees that the electron spin operators are
cancelled in all the odd terms of the ME ($k$ odd), at the expense
of making $T_c$ twice as long as in PDD.  As long as the bath is
non-dynamical, $H_B=0$, this also yields $\overline {\cal
H}^{(1)}=0$. As one may verify by direct calculation, the
lowest-order term containing the electron spin is
\begin{eqnarray}
\overline {\cal H}^{(2)} &\hspace*{-2mm}=\hspace*{-2mm}
& -{\frac{\tau^3}{6T_c}}\sum_{k=1}^{8}
\sum_{j=1}^{k}\sum_{i=1}^{j} \hspace*{-0.8mm}
\Big(\left[H_k,[H_j, H_i]\right]
+ \left[[H_k,H_j], H_i\right]\Big) \nonumber \\
&\hspace*{-2mm}\propto \hspace*{-1.5mm}& S_z {{\tau}^{2}}.
\label{eq:sddh2}
\end{eqnarray}

A powerful method to enhance DD performance by
systematically shrinking the norm of higher-order terms is to resort
to concatenated design \cite{Khodjasteh05}.  CDD relies on a temporal
recursive structure, so that at level $\ell+1$ the protocol is defined
recursively as $C_{\ell+1}=C_{\ell}XC_{\ell}ZC_{\ell}XC_{\ell}Z$,
$C_0$ denoting as before free evolution under $H$.  Building on the
results of Ref.~\onlinecite{Zhang07a}, we focus in this work on a
{\em truncated}
version of CDD, whereby concatenation is stopped at a certain level
and the resulting supercycle is repeated periodically afterwards.
Such a protocol is referred to as PCDD$_\ell$, with
$T_c=4^{\ell}\tau$ (note that $\ell=1$ recovers PDD).

Specifically, a cycle of the PCDD$_2$ protocol consists of two
identical half-cycles, and has a form
\begin{equation}
[C_0 X C_0 Z C_0 X C_0 Y C_0 X C_0 Z C_0 X C_0][\sf repeat],
\label{cddcycle}
\end{equation}
\noindent where $[\cdot]$ denotes each half-cycle. This protocol
leads to especially remarkable DD performance and averaging
structure for the central-spin problem. First, because the
protocol is time-symmetric, averaging is at least of second order,
as in SDD that is, $\overline {\cal H}^{(0)} = 0$, and $\overline
{\cal H}^{(1)} = 0$.  However, in contrast to SDD, the
second-order contribution $\overline {\cal H}^{(2)}$ turns out to
be an effective {\em pure bath} term which renormalizes the bath
Hamiltonian without directly contributing to the decoherence
dynamics.  Thus, operators mixing electron and nuclear spins can
only appear at order $\overline {\cal H}^{(4)}$ and higher. The
fact that a pure-bath contribution $\overline {\cal H}^{(2)}$ is
generated to second-order in the ME, and that a particularly
favorable scenario is to be expected for CDD convergence, was
implied in Ref.~\onlinecite{Khodjasteh06}.  Remarkably, two
additional features emerge to second order in the controlled
dynamics:

(i) The DD-induced bath Hamiltonian has a regular coupling structure,
\begin{eqnarray}
\overline {\cal H}^{(2)} &\hspace*{-2mm}=\hspace*{-2mm}
& -{\frac{\tau^2}{96}}\sum_{k=1}^{16}
\sum_{j=1}^{k}\sum_{i=1}^{j}
\Big(\left[H_k,[H_j, H_i]\right]
+ \left[[H_k,H_j], H_i\right]\Big)
 \nonumber \\
&\hspace*{-2mm}= \hspace*{-2mm}&
 \sum_{i=1}^N \sum_{j<i=1}^N {\Gamma'}_{i j}
\Big({\mathbf I}_i \cdot {\mathbf I}_j - 3 I_i^x I_j^x \Big),
\label{renHB}
\end{eqnarray}
where
\begin{equation}
{\Gamma'}_{i j}=\tau^2 {\frac{A_i A_j (A_i+A_j)}{3}}.
\label{gammap}
\end{equation}
\noindent
That is, to second order in $\tau$, the hyperfine interaction between
the electron and the nuclei is removed, and an effective dipolar
Hamiltonian with control-renormalized couplings is induced on the
nuclear spins -- compare with the Hamiltonian in
Eq.~(\ref{Hb})~\cite{Xremark}.

(ii) To the same level of accuracy in $\tau$, it is possible to show
that a similar averaging is achieved by a {\em half-PCCD}$_2$ protocol
-- that is, a protocol whose cycle consists of just the first half in
(\ref{cddcycle}). In practice, this may be useful to reduce the number
of required pulses for a given desired accuracy.

\subsubsection{Randomized DD}

Randomized design offers another approach to improve DD performance,
by both ensuring robust behavior in the presence of intrinsically
time-varying open-system Hamiltonians and by minimizing the impact of
coherent error accumulation at long evolution times
\cite{Viola05,Kern05,Santos06,Santos06b}.  Unlike deterministic
schemes, randomized DD is distinguished by the fact that the future
control path is not fully known in advance.  Analysis is most
conveniently carried out in a logical frame that follows the applied
control~\cite{Viola05}, and by tracking the applied sequence to ensure
that appropriate frame compensation can be used to infer the evolution
in the physical frame.  Although loss of periodicity in the control
Hamiltonian causes AHT not to be directly applicable, contributions of
different orders in $\tau$ may still be identified in the effective
Hamiltonian describing a given control sequence.

Among all randomized protocols \cite{Kern05,Santos06,Santos06b}, we
consider a few representative choices. {\em Na{i}ve random DD} (NRD)
corresponds to changing the control propagator according to a path
which is uniformly random with respect to the invariant Haar measure
on ${\cal G}$. For our system, this means that a $X$, $Y$, $Z$ pulse,
or a no-pulse is applied with equal probability at every instant
$t_j=j\tau$, $j=0,1,\dots$. So-called {\em random path DD} (RPD)
merges features from both pure random and cyclic design, by involving PDD cycles where the path to traverse ${\cal G}$ is each time chosen
at random among the $|{\cal G}|!$ possible ones.  For our system, we implement a simplified pseudo-RPD protocol by restricting to cycles which always begin with the identity -- that is, at every instant
$T_n=4n\tau$, $n=1,2,\ldots$, we randomly choose a
sequence from the following set of $(|{\cal G}|-1)!$ possibilities:
\begin{eqnarray}
\{C_0XC_0ZC_0XC_0Z,\: C_0ZC_0XC_0ZC_0X, \nonumber \\
C_0XC_0YC_0XC_0Y,\: C_0YC_0XC_0YC_0X, \nonumber \\
C_0YC_0ZC_0YC_0Z,\: C_0ZC_0YC_0ZC_0Y\}. \nonumber
\end{eqnarray}
Unlike NRD, RPD guarantees that averaging of $H_{SB}$ is retained to
lowest order over each control cycle. By enforcing symmetrization on
the RPD protocol, {\em symmetrized random path DD} (SRPD) is obtained,
whereby every sequence from the above set is augmented by its
time-reversed counterpart. SRPD additionally ensures that all odd
terms in the effective Hamiltonian which involve the electron spin
operators disappear at $T_p=8p\tau$, $p=1,2,\ldots$.

\subsection{DD performance metrics}

In order to meaningfully compare different DD protocols for given
control resources (in particular, finite $\tau$), an appropriate
control metric should be identified. Different choices may best
suit different control scenarios.
Quantities such as pure-state input-output fidelity and purity, for
instance, depend strongly on the initial state of the central system,
thus being appropriate when preservation of a {\em known} electron initial
state is the intended DD goal.  Average input-output fidelity and gate
entanglement fidelity do not rely on a particular initial state,
however they depend on the probability distribution of the initial
states and quantify typical
performance~\cite{Schumacher96,Nielsen02,Fortunato02}.
Taking advantage of the small dimensionality of the central spin
system, an accurate way for quantifying DD performance at
preserving an {\em arbitrary} electron pure state is to evaluate
both best-case and worst-case input-output
fidelities.

For a given initial state $|\psi_S(0) \rangle$, recall that the
input-output fidelity is defined as
\begin{eqnarray}
F(t)={\rm Tr}[\rho_S (t)\rho_S (0)]
=\langle \psi_S(0) |\rho_S(t)|\psi_S(0) \rangle,
\label{fide}
\end{eqnarray}
where $\rho_S(t)$ is the reduced density matrix of the central spin
$S$ at time $t$, $\rho_S(t)={\rm Tr}_B \rho(t)$,
$\rho(t)$ and ${\rm Tr}_B$ denoting the total density
operator at time $t$, and the partial trace over the bath degrees
of freedom, respectively.  $F(t)$ gives a measure of how far the
central system has evolved away from its initial state. The
best-case ($b$) and the worst-case ($w$) fidelities are then
naturally defined as
\begin{eqnarray*}
F_{b}(t)=\mbox{max}_{|\psi_S \rangle} \{F(t)\}, \;\; \;
F_{w}(t)=\mbox{min}_{|\psi_S \rangle} \{F(t)\}.
\end{eqnarray*}

\subsection{Numerical methodology}
\label{sec:Num}

Analytical bounds on the expected worst-case fidelity decay for
various DD protocols have been obtained for sufficiently short
evolution times based on either AHT and/or on additional
simplifying assumptions~\cite{Viola05,Kern05,
Santos06,Khodjasteh05,Santos06b,Khodjasteh06}.  For cyclic schemes,
the relevant error bounds assume convergence of the ME series, thus
requiring, in particular, that $\omega_c T _c < 1$,
where in our problem $\omega_c$ is the highest frequency of the total
(electron plus nuclei) spectrum. For a typical GaAs QD with $N\sim
10^6$ bath spins,
$$\omega_c\approx \sum_k \frac{|A_k|}{4} \sim \frac{N A}{4}
\sim 20~\mbox{GHz},$$
\noindent
thus strict convergence of the ME implies extremely short
characteristic control time scales, $\tau \sim 10$~ps.
Moreover, for ESR experiments under resonance conditions,
$\omega_c\sim 20$~GHz is about $400$ times larger than
currently available carrier pulse
frequency \cite{Rimberg06}, and roughly $20$ times larger
than attainable exchange-gating frequency~\cite{Coish07, Petta05}.

In order to evade the strict convergence requirement and to study DD
performance in regimes of more direct experimental relevance,
numerical simulations are necessary.  Specifically, we are interested
in pulse separations of the order of $1/\sigma$, where
\begin{equation}
\sigma=\frac{1}{2}\sqrt{\sum_k A_k^2}=
\frac{1}{2} \sqrt{N} A
\label{sigma}
\end{equation}
is the characteristic {\em half-width} of the coupling spectrum, as
opposed to the highest frequency $\omega_c$, with $\sigma$ being
roughly $\sqrt{N}\sim 10^3$ times smaller than $\omega_c$.  Thus,
$T_c\ge 4\tau$ $\sim \sqrt{N}\omega_c^{-1}$, with typical inter-pulse
delays values of the order of $10$ ns, which is not too far (within an
order of magnitude) from current experimental capabilities
\cite{Rimberg06}. Furthermore, in order to access DD in the long-time
limit, we consider up to thousands of control cycles.  Nuclear spin
environments consisting of $N \leq 25$ spin-$1/2$ are investigated
(with a corresponding Hilbert space dimension of $\sim 7\cdot 10^7$)
-- giving hope that our main conclusions may be extrapolated to real
mesoscopic environments.

The initial state of the entire system is taken to be a direct product
of the electron initial spin state $\rho_S(0)=|\psi_S (0) \rangle
\langle \psi_S (0)|$ and the bath initial spin state $\rho_B(0)$.  In
most cases, we assume an {\em unpolarized} spin bath, described by the
thermal equilibrium density operator $\rho_B(0)= 2^{-N} I_{2^N\times
2^N}$, where $I_{2^N\times 2^N}$ is the $2^N$-dimensional identity
matrix (polarized initial bath states shall be considered in
Sec.~\ref{sec:bp}).  Such a maximally mixed spin state reflects the
fact that for typical experimental dilution-refrigerator temperatures
(T $\sim 100$ mK), the thermal energy is much larger than the energy
scale of the intra-bath spin interactions. For a small number of bath
spins ($N\le 8$), we directly simulate the evolution of the total
system's density operator, followed by a partial trace over $B$. For
larger $N$, this approach is not computationally efficient, and we
perform simulations by assuming that the total system is in a pure
state~\cite{Zhang06,Zhang07r,Zhang07a,Zhang07b}. In this case,
$\rho_B(0)$ is approximated as a uniformly random superposition of
environment product states, $|\psi_B(0)\rangle = \sum_{i=1}^{2^N} c_i
|\phi_i \rangle$ of all possible tensor products of the form $|\phi
\rangle=|\uparrow\rangle_1 \otimes |\downarrow\rangle_2 \otimes
|\downarrow\rangle_3 \otimes \ldots \otimes |\uparrow\rangle_N$, where
$c_i$ are uniformly distributed random
numbers~\cite{Zhang06,Zhang07r,Popescu05,Zhang07a,Zhang07b},
subject to $\sum_i |c_i|^2=1$.
Thanks to concentration of measure on the space of random pure states,
such an approximation has an exponentially small error, of the order
of $2^{-N/2}$, with respect to using the identity state -- that is,
$\sim 0.5$\% for $N=15$.

The desired best/worst case fidelity $F_{b/w}$ is evaluated by
invoking quantum process tomography~\cite{Nielsen00}.  Four different
initial states of the electron are considered,
\begin{eqnarray*}
|\psi_z\rangle &=& |\uparrow \rangle, \\
|\psi_{\bar z}\rangle &=& |\downarrow \rangle, \\
|\psi_x \rangle &=& \frac{1}{\sqrt{2}}
\left(|\uparrow \rangle + |\downarrow\rangle\right),\\
|\psi_{y}\rangle &=& \frac{1}{\sqrt{2}} \left(|\uparrow \rangle +
i|\downarrow\rangle\right)
\end{eqnarray*}
and for each such state the time-dependent Schr\"odinger equation
of the total system subjected to DD is solved, by employing the
Chebyshev polynomial expansion method to calculate the evolution
operator~\cite{Dobrovitski03, Zhang07r}. At the final evolution
time $T$, the four reduced density matrices (obtained upon partial
trace over $B$) are used to compute the superoperator matrix
$\chi_{mn} (T)$ which describes the electron spin dynamics
according to the linear map
\begin{eqnarray*}
\tilde{\rho}_S(T) = \sum_{m=0}^3\sum_{n=0}^3 K_m \rho_S(0)
K_n^\dag \, \chi_{mn}(T)\,,
\end{eqnarray*}
where $K_0=I$, $K_1=X$, $ K_2=-iY$, $K_3=Z$, and $\tilde{\rho}_S$
specifies evolution in the logical frame, with
$\tilde{\rho}_S(0)={\rho}_S(0)$.  Since an arbitrary initial pure
state of $S$ on the Bloch sphere may be parameterized as
$$
|\psi_S(0)\rangle = \cos(\theta/2) |\uparrow \rangle
+\sin(\theta/2)e^{i\varphi}|\downarrow \rangle,
$$
with $\theta \in [0,\pi]$ and $\varphi \in [0,2\pi]$,
$F_{b/w}(T)$ simplifies to
$$
F_{b/w}(T)=\mbox{max/min}_{(\theta, \varphi)} \,\langle
\psi_S(0)|\tilde{\rho}_S (T)|\psi_S(0)\rangle \,.
$$
In practice, after determining the matrix $\chi$, we obtain
$F_{b/w}(T)$ by numerical optimization, using a statistically
meaningful set of initial guesses for $(\theta,\varphi)$.


\section{Results on DD performance}
\label{sec:performance}

In this section we summarize results on best- and worst-case
performance of deterministic DD protocols relevant to the QD problem,
including analytical estimates at short times. The discussion of
randomized protocols is postponed to Sec.~\ref{sec:fw}, as there is
little difference between best and worst case.

\subsection{Best-case performance and decoherence freezing}
\label{sec:fb}

Investigation of best-case performance both provides an upper bound on
the attainable DD fidelity, and reveals a remarkable phenomenon of
single-qubit deterministic DD: the optimal performance bound is
reached for initial states of the electron, which depend on the
geometry of the control protocol and cause the long-time fidelity to
freeze at a non-zero value determined by the DD rate
\cite{Zhang07a,Zhang07b}.

\subsubsection{Short-time behavior}
\label{fb1}

In order to bridge analytical and numerical results, we begin
by considering the performance of a single DD cycle, in the
limit where the inter-pulse delay $\tau$ is small enough to justify
the application of AHT and the ME.  Clearly, perfect (stroboscopic)
preservation would be ensured for an initial electron spin state which
is an eigenstate of the full AHT $\overline{\cal H}$.  For finite DD
accuracy, the best-case scenario still corresponds to initializing the
electron in an eigenstate of the lowest-order term of the ME, so that
the fidelity decay after one cycle is determined by the
next-lowest-order contribution to $\overline{\cal H}$.

Let the total cycle propagator be expanded as
\begin{eqnarray}
U (T_c) &=& \exp[-iT_c(\overline{\cal H}^{(0)}+\overline{\cal
H}^{(1)}+\overline{\cal H}^{(2)}+...)],\;\;\;
\label{Ucycle}
\end{eqnarray}
and consider, to begin with, the simplest $[C_0ZC_0Z]$ CPMG
protocol. Then $\overline{\cal H}^{(0)}= S_z \otimes
\sum_k A_k I_k^z$, and the
next dominant contribution arises from $\overline{\cal H}^{(1)}$. By
using Eq.~(\ref{fide}), and by assuming that $|\psi_S(0)\rangle$ is
invariant under $\overline{\cal H}^{(0)}$ for best-case performance,
one finds
\begin{eqnarray}
F_b (T_c)^{\mbox{\small CPMG}} &\approx & 1- T_c^2 \mbox{Tr}_B \Big[
\rho_B(0) \left(\Delta \overline{\cal H}^{(1)}
\right)^2_{|\psi_S(0)\rangle} \Big]
\nonumber \\
&\approx & 1-\alpha T_c^2 \tau^2 + {\mathcal O}(T_c^4 \tau^2),
\end{eqnarray}
where
$$\Big(\Delta \overline{\cal H}^{(k)}\Big)^2_{|\psi_S\rangle}=
\langle \psi_S | (\overline{\cal H}^{(k)})^2
|\psi_S \rangle - \langle \psi_S | \overline{\cal H}^{(k)}
|\psi_S \rangle ^2$$
\noindent
denotes the partial variance of the dominant correction
$\overline{\cal H}^{(k)}$ in the initial electron state
$|\psi_S\rangle$, and $\alpha \tau^2$, $\alpha \in {\mathbb R}$,
gives the expectation of such variance operator over the
initial bath state.  Thus, $\alpha$ is clearly protocol-
and bath-state dependent. By a similar calculation, the
best-case fidelities for PDD and SDD may be expressed as:
\begin{eqnarray}
F_b (T_c)^{\mbox{\small PDD}} &\approx &
1- T_c^2 \mbox{Tr}_B \Big[
\rho_B(0) \left(\Delta \overline{\cal H}^{(2)}
\right)^2_{|\psi_S(0)\rangle} \Big]
\nonumber \\
&\approx & 1-\beta T_c^2 \tau^4 + {\mathcal O}(T_c^4 \tau^6),
\end{eqnarray}
\begin{eqnarray}
F_b (T_c)^{\mbox{\small SDD}} &\approx &
1- T_c^2 \mbox{Tr}_B \Big[
\rho_B(0) \left(\Delta \overline{\cal H}^{(4)}
\right)^2_{|\psi_S(0)\rangle} \Big]
\nonumber \\
&\approx & 1-\gamma\, T_c^2 \tau^8 + {\mathcal O}(T_c^4 \tau^{12}),
\end{eqnarray}
for suitable real parameters $\beta, \gamma$. Thus, $k=2, 4$ for
PDD and SDD, respectively.

Beside the above analytical determination, the relevant fidelity
decay terms may also be obtained symbolically -- by Taylor-expanding
the exact evolution operator $U(T_c)=\prod_j e^{-i H_j \tau}$ for
the appropriate transformed Hamiltonians $H_j$, and collecting
coefficients of order $\tau^n$, so that the lowest two
contributions to $\overline{\cal H}$ may be isolated. This
approach is required, in particular, to extract the best-case
leading decoherence order for PCDD$_2$, which we could not
calculate analytically \cite{Nremark}.  The single-cycle results are
presented in Table~\ref{table:sb}; an excellent agreement is seen
between analytical and symbolic results, as reported in the second
and third row, respectively (see also Appendix~\ref{sec:cpmg} for yet
another analytical derivation of the coefficient $\alpha$ and the
leading decoherence order $n$ for CPMG, which agrees well with the
results given by the other methods).

\begin{table}[b]
\caption{Fitting parameters of the best-case cycle performance $F_b (T_c)=
1-\kappa\,(\omega_c\tau)^n T_c^2$, $\kappa \in {\mathbb R}$, for a
single DD cycle under CPMG, PDD, SDD, and PCDD$_2$ in the small
$\tau$ region where $\omega_c T_c \lesssim 1$. Bath size $N=5$.
Also presented are ME predictions by analytical method,
$n({\rm ME})$, and symbolic Taylor expansion, $n({\rm Sym})$.
The same agreement holds for $N=3$ and $N=7$\vspace*{3mm}.}
\begin{tabular}{c|cccc}\hline\hline
& CPMG & PDD & SDD & PCDD$_2$ \\
\hline
$n({\rm Fit})$ &2.00&4.02&7.93&11.80 \\
$n({\rm ME})$&2&4&8&-\\
$n({\rm Sym})$&2&4&8&12\\
\hline\hline
\end{tabular}
\label{table:sb}    
\end{table}

\begin{figure}[h]
\includegraphics[width=3in]{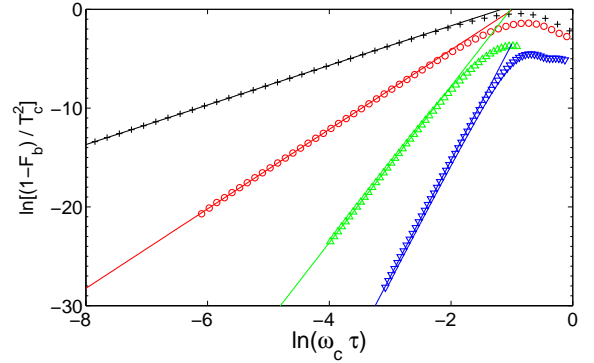} 
\caption{(Color online) Best-case performance for single-cycle DD
under the CPMG (plus signs), PDD (circles), SDD (upward pointing
triangles), and PCDD$_2$ (downward pointing triangles) protocol in
the small $\tau$ region. Bath size $N=5$. Solid lines are linear
fits, with fitting parameters given in Tab.~\ref{table:sb}.}
\label{fig:fsb}
\end{figure}

An alternative way for determining $F_b(T_c)$ is provided by direct
numerical simulation of the total evolution, followed by the process
tomography procedure described in Sec. \ref{sec:Num}. By restricting the
simulations to short $\tau$, and fitting $1-F_b(T_c)$ to a power-law
function of $\tau$, it is then possible to independently determine the
leading decoherence orders in the ME series.  Fig.~\ref{fig:fsb}, for
instance, shows the dependence upon $\tau$ of $F_b(T_c)$ for
$N=5$. The initial region in Fig.~\ref{fig:fsb}, where $1-F_b(T_c)$
has a power-law dependence on $\tau$, indicates the region of validity
for AHT/ME, the slope of the curve determining the order of the
best-case decoherence term.  The first line of Table~\ref{table:sb}
shows the leading ME order, $n$(ME), obtained in this way, which is in
excellent agreement with the available analytical estimates obtained
from AHT/ME. In particular, these simulations confirm that
$\overline{\cal H}^{(6)}$ is the leading decoherence term for
PCDD$_2$ in the best-case scenario.

Table~\ref{table:sb} and Fig.~\ref{fig:fsb} clearly demonstrate
how DD performance improves as we go from PDD to SDD and to
PCDD$_2$. Even though, as already remarked, the CPMG protocol
$[C_0ZC_0 Z]$ does not achieve maximal DD for zero bias field, the
existence of an approximate integral of motion, $S_z$, still makes
it possible to decouple with high fidelity {\em provided} that the
initial electron spin state is a $S_z$-eigenstate. For two-axis
cyclic DD, a preferred direction for initialization may still be
identified. The PDD protocol $[C_0XC_0 ZC_0 XC_0 Z ]$, for
instance, conserves $S_y$ in the limit $\tau\to 0$ (because
$Z\,X\sim Y$, which coincides with the half-cycle direction); this
establishes a $S_y$-eigenstate as the best-case state for PDD.
When several DD cycles are implemented, the existence of
approximate control-induced symmetries is responsible for the
decoherence freezing phenomenon which we address next.

\subsubsection{Long-time behavior}

At long times, none of the above-mentioned analytical or symbolic
methods is applicable, thus numerical simulation is required.

\begin{figure}
\includegraphics[width=3.25in]{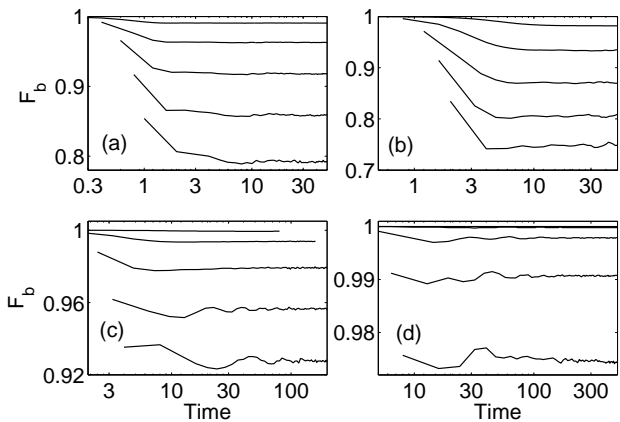}
\caption{Best-case performance of (a) CPMG, (b) PDD, (c) SDD, and
(d) PCDD$_2$ starting from an initial state along the half-cycle
direction ($z$ for (a) and (c); $y$ for (b) and (d), respectively).
Bath size: $N=15$. $\tau=0.1$, $0.2$,
$0.3$, $0.4$, $0.5$, from top to bottom in all panels (in panel (d),
the curves for $\tau=0.1$ and $0.2$ are indistinguishable).
Decoherence freezes at sufficiently long time.}
\label{fig:fb}
\end{figure}

Fig.~\ref{fig:fb} shows the long-time best-case performance for
the deterministic control protocols discussed so far.
{\em Decoherence freezing} is clearly seen as a plateau of $F_b$ at
sufficiently long evolution times~\cite{Viola98e,Zhang07a}, the
corresponding saturation value increasing as the pulse delay $\tau$
decreases. From a control standpoint, decoherence freezing may be
thought of as signaling the dynamical generation of a stable
one-dimensional decoherence free subspace via DD~\cite{Viola00,
Wu02}. In NMR language, the resulting saturation is reminiscent of the
``pedestals'' seen in the long-time magnetization signal under pulsed
spin-locking conditions~\cite{Haeberlen76, Suwelack80,
Ladd05}. Physically, one may think that an {\em effective magnetic
field} is created by the DD pulses, and that proper alignment of the
initial spin prevents the electron from precessing around this
direction -- thereby experiencing fully frozen nuclear spin
fluctuations.  (See also Ref.~\onlinecite{Oulton07} for a recent
demonstration of a similar-in-spirit stabilization effect at zero
applied field, leading to a long-lived nuclear-spin polaron state via
optical pumping with polarized light).
In practice, decoherence freezing may be exploited to optimally
preserve a {\em known} initial electron spin state, through
appropriate design of a DD protocol with the desired quasi-integral of
motion.  While a similar effect could be achieved by applying a strong
static bias field, one advantage of DD stabilization is that better
storage may be ensured by simply rearranging the pulse sequence, so
that it implements a higher-level protocol (from PDD to PCDD$_2$, for
example).

\begin{figure}[t]
\includegraphics[width=3in]{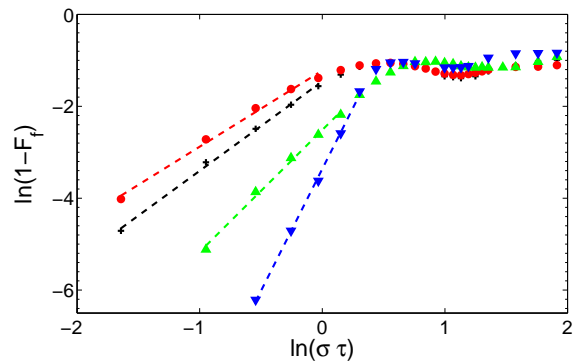}
\caption{(Color online) Dependence of the coherence saturation value
on pulse delay for CPMG (plus signs), PDD (circles), SDD (upward pointing
triangles), and PCDD$_2$ (downward pointing triangles). Dashed
lines are linear fittings at small $\tau$.} \label{fig:fbf}
\end{figure}

While asymptotic saturation behavior has been reported for purely
dephasing spin-boson models with arbitrary initial spin
states~\cite{Viola98, Viola99, Viola98e, Shiokawa04, Santos05},
saturation effects for a spin bath have only received attention very
recently~\cite{Yao07, Zhang07a, Zhang07b}. Thanks to its inherent
simplicity, the CPMG protocol makes it possible to gain analytical
insight under the assumption that a simplified coupling Hamiltonian is
appropriate:
\begin{eqnarray}
H^{QSA}_{SB} &=& A {\mathbf S}\cdot \sum_{k=1}^N {\mathbf I}_k.
\label{eq:ha}
\end{eqnarray}
This corresponds to assuming a uniform electron density in the QD,
allowing one to regard the total nuclear spin as a constant. For our
purposes, because of the large number of spins in the bath, this is
also equivalent to describing the nuclear spin reservoir under the
so-called Quasi-Static Approximation~(QSA), which treats the
Overhauser nuclear field as a classical random static
field~\cite{Taylor06}.  While in principle the QSA is valid for short
evolution times~\cite{Taylor06,Zhang06}, it is important to realize
that the domain of validity of the QSA is extended in the presence of
DD pulses, similarly to what happens for an external bias
field~\cite{Taylor06}, and consistent with the fact that the nuclear
field becomes progressively more static relative to the electron
dynamics as the electron-nuclear coupling is suppressed.

By following the steps illustrated in Appendix~\ref{sec:cpmg}, the
freezing value reported in Ref. \onlinecite{Zhang07a} is obtained:
$$F_{f}^{\mbox{\small CPMG}}
\approx 1-\frac{\tau^2}{2T_2^{*2}},$$
\noindent
for sufficiently small $\tau$. Interestingly, the leading power
of $\tau$, $n=2$, does not explicitly depend on the number $N$
of bath spins.  For other protocols in the same range of $\tau$,
numerical results (see
Fig.~\ref{fig:fbf}) suggest a similar dependence of the asymptotic
coherence value,
$$F_f\approx 1- a (\sigma \tau)^n, \;\;\; a \in {\mathbb R},$$
\noindent
with the relevant values of $n$ being given in
Table~\ref{table:sat}. Note that the characteristic values
of $\tau$ considered here are of order of $1/\sigma$, that is,
$\omega_c \tau\gg 1$ in the simulations, thereby well beyond
the convergence region of AHT/ME.

\begin{table}[h]
\caption{Fitting parameters of the decoherence freezing value $F_f =
1-a (\sigma\tau)^n$ for CPMG, PDD, SDD, and PCDD$_2$ at small
$\tau$, where $\sigma\tau \lesssim 1$ but
\vspace*{3mm}
$\omega_c\tau\gg 1$.}
\begin{tabular}{c|cccc}
\hline\hline
& CPMG & PDD & SDD & PCDD$_2$ \\
\hline
$n({\rm Fit})$ & 2.0 & 1.7 & 2.7 & 5.3 \\
\hline\hline
\end{tabular}
\label{table:sat}
\end{table}


\subsection{Worst-case performance and arbitrary state preservation}
\label{sec:fw}

As evidenced by the best-case considerations presented above, for
sufficiently small $\tau$, initial electron spin states which are
(approximate) eigenstates of the decoupled evolution are stable at
long times, whereas the spin components perpendicular to the
decoherence-free axis are lost in the long-time regime. In such a
picture, the worst case scenario for a given cyclic protocol
corresponds to initial spin states which are perpendicular to the
effective half-cycle control axis.  Clearly, worst-case performance
lower-bounds the fidelity of storage achievable for an {\em arbitrary}
(possibly unknown) initial state, as required for an electron-spin
quantum memory.

\subsubsection{Short-time behavior}

\begin{figure}[t]
\includegraphics[width=3in]{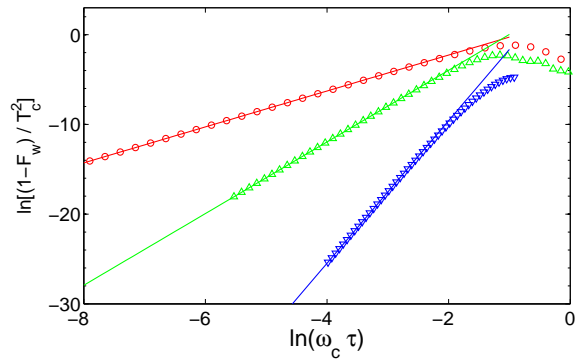} 
\caption{(Color online) Worst-case single-cycle performance for
PDD (circles), SDD (upward pointing triangles), and PCDD$_2$
(downward pointing triangles) in the small $\tau$ region. Bath
size: $N=5$.  Solid lines are linear fittings, with fitting
parameters given in Tab.~\ref{table:sw}.} \label{fig:fsw}
\end{figure}

In order to quantitatively assess worst-case DD fidelities, we begin,
as in the previous Section, by examining single-cycle
performance. Again, the order of the leading decoherence term is
determined using three methods: (i) exact numerical simulation;
(ii) analytical predictions based on AHT/ME; and (iii) symbolic
Taylor expansion.  Fig.~\ref{fig:fsw} shows the exact
$\tau$-dependence of the worst-case fidelity for single-cycle DD.
Similar to the best case, $F_w (T_c)$ depends on
$\tau$ according to a power-law, when $\tau$ is sufficiently small to
ensure that $\omega_c T_c \lesssim 1$ and that the AHT/ME approach is
valid. Table~\ref{table:sw} (first line) gives the leading decoherence
term orders extracted from Fig.~\ref{fig:fsw}, which are in excellent
agreement with the analytical and symbolic predictions (last two lines).
Note that CPMG, which is not a universal decoupling sequence, is not useful in the worst-case scenario, thus we do not consider it further.

\begin{table}[h]
\caption{Fitting parameters of worst-case single-cycle performance
$F_w (T_c) =1-\kappa (\omega_c\tau)^n T_c^2$, $\kappa \in {\mathbb R}$,
for PDD, SDD, and PCDD$_2$ in the small $\tau$ \vspace*{3mm} region
where $\omega_c \tau \lesssim 1$.}
\begin{tabular}{c|cccc}
\hline\hline
& PDD & SDD & PCDD$_2$ \\
\hline
$n({\rm Fit})$&2.00&3.99&7.94 \\
$n$(ME)&2&4&8\\
$n$(Sym)&2&4&8 \\
\hline\hline
\end{tabular}
\label{table:sw}    
\end{table}

\subsubsection{Long-time behavior and randomized DD}

\begin{figure}[t]
\includegraphics[width=3.4in]{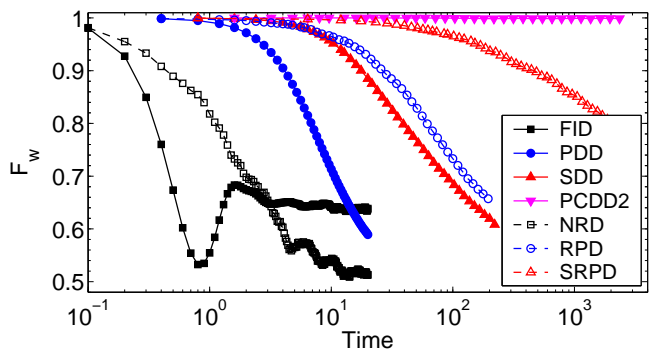}
\caption{(Color online) Worst-case DD performance in the logical
frame~\cite{Viola05, Santos06b}, with $\tau=0.1$. Hamiltonian parameters
are: $\omega_0=0$, $\Gamma_0=0$ and $N=15$. For deterministic DD, data
points are acquired at the completion of each cycle, while for NRD and
FID this is done after every $\tau$, for RPD after every $4\tau$, and
for SRPD after every $8\tau$.  Random protocols are averaged over 100
control realizations.}
\label{fig:cp}
\end{figure}

For long evolution times, the worst-case performance $F_w(T)$ of
six DD protocols (three deterministic and three randomized) are
summarized in Fig.~\ref{fig:cp}.  All schemes lead to substantial
enhancement of the electron spin coherence, some of them by more
than a factor of 1000, with PCDD$_2$ showing the most dramatic
improvement \cite{Zhang07a}. According to the ME, the leading
order term for the FID signal is $\overline{\cal H}^{(0)}$,
whereas it is $\overline{\cal H}^{(1)}$ for PDD, $\overline{\cal
H}^{(2)}$ for SDD, and $\overline{\cal H}^{(4)}$ for PCDD$_2$, as
discussed in Sec. \ref{twoaxis}.  For randomized protocols, lack
of periodicity prevents the definition of a time-independent
average Hamiltonian, thus AHT is not applicable.  However, for a
given evolution time, an effective Hamiltonian and the
corresponding leading orders to coherence decay may still be
defined directly in terms of the unitary logical-frame propagator
\cite{Santos06,Santos06b}.

In general, protocols with higher leading order tend to give superior
performance.  However, such a conclusion does {\em not} necessarily
hold if a randomized protocol is compared to a deterministic one.
For example, RPD has a lower-order leading term than SDD,
so that SDD may outperform RPD at short times. However,
Fig.~\ref{fig:cp} shows that at long times RPD outperforms SDD, which
demonstrates the advantage of randomization in suppressing coherent
error accumulation. The poor performance of NRD is expected, since the
advantages of this low-level DD scheme may emerge only when ${\cal G}$
is large, and cyclic DD is inefficient. In a closed
system~\cite{Santos06}, SRPD has been found superior to PCDD$_2$ at
long times, but for the QD model considered here SRPD does not match
PCDD$_2$, confirming the fact that irreducible DD groups and slow
baths are especially favorable for concatenated
control~\cite{Khodjasteh05}.  However, it is worth emphasizing that
increase in the concatenation level for fixed pulse separation does
{\em not} necessarily improve the protocol performance: As shown in
Ref.~\onlinecite{Zhang07a}, PCDD$_4$ may deliver worse fidelity than
PCDD$_2$ if $\tau$ becomes sufficiently large.  A possible explanation
may be rooted in the DD-induced renormalization of the pure-bath terms
discussed in Sec. \ref{twoaxis}, which for large $\tau$ may become
important enough to offset the benefits associated with a more
elaborated DD cycle.

\section{Real-system considerations}
\label{sec:real}

In realistic scenarios, several factors beyond the simplified
treatment considered thus far will unavoidably affect DD
performance.  Among those related to the underlying QD model
Hamiltonian, the most important are the influence of spin-bath
dynamics, external bias fields, and nuclear spin polarization.
We shall address these factors one by one, by primarily focusing
on the worst-case fidelity of the PCDD$_2$ protocol, which has
emerged as the best performer for the problem under exam.

\subsection{Bath size}

\begin{figure}[b]
\includegraphics[width=3.2in]{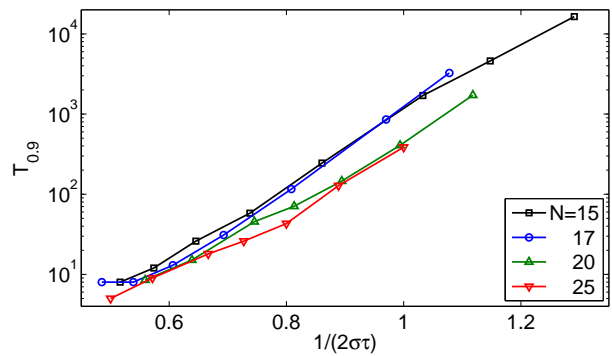}
\caption{(Color online) Bath size effect. $T_{0.9}$ vs.
$1/(2\sigma\tau)$ for PCDD$_2$ with different bath sizes $N=15$
(black squares), $17$ (blue circles), 20 (green upward pointing
triangles), and $25$ (red downward pointing triangles).}
\label{fig:size}
\end{figure}

In our simulations, the number of bath spins is moderately large,
$N\le 25$, thus it is essential to assess to what extent our numerical
results might be applicable to real QD devices with $N\sim
10^4$--$10^6$.  In order to verify this, PCDD$_2$ simulations have
been carried out for baths with $N$ varying from $15$ to $25$, their
corresponding spectral width being characterized by $\sigma=
\sqrt{N}A/2$ as in Eq.~(\ref{sigma}).  Fig.~\ref{fig:size} illustrates,
for different $N$, the instant of time $T_{0.9}$ where $F_w(T)$ for
PCDD$_2$ reaches a threshold value of $0.9$ -- as a function of the
dimensionless parameter $1/(2 \sigma \tau)$.  It is seen how, by
correctly rescaling $\tau$ (that is, {\em by measuring $\tau$ in units
of $1/\sigma$}), the curves corresponding to different bath size tend
to fall on top of each other, especially as $N$ increases.  This provides
strong evidence that our results should be applicable to realistic
mesoscopic spin environments upon appropriate parameter scaling.

It is also interesting to stress that $T_{0.9}$ increases
extremely rapidly as the product $\sigma\tau$ decreases and the
region of very fast DD is entered (note the logarithmic scale of
the $y$-axis in Fig.~\ref{fig:size}). This rapid increase has been
analyzed earlier~\cite{Zhang07b}.  We remark here that a na{i}ve
extrapolation from the ME, $T_{0.9}\propto 1/\tau^4$, which could
be expected to hold in the limit $\tau \rightarrow 0$, strongly
disagrees with our data in the long-time parameter range of
Fig.~\ref{fig:size}, and that a Zeno-type analysis as invoked in
Ref.~\onlinecite{Zhang07b} appears more appropriate to explain the
observed $\tau$-dependence.

\subsection{Intrabath interaction}

The effect of the internal bath Hamiltonian $H_B$, Eq.~(\ref{Hb}), may
become important once the electron coherence time becomes longer than
the characteristic time scale of the corresponding bath evolution. As
remarked, this effect may be enhanced in principle by the
PCDD$_2$-induced pure-bath dipolar contribution,
given in Eq.~(\ref{renHB}). In order to assess at which point
DD performance start to be significantly affected by nuclear
spin dynamics, we have performed numerical simulations by
choosing $\Gamma_{kl}$ as uniformly random numbers in
$[-\Gamma_0, \Gamma_0]$, and by manually increasing $\Gamma_0$ up to
values comparable to $0.1 A_k$, to keep the simulation time
reasonably short.  In this way, in view of Eq. (\ref{gammap}),
we cause the effects of $H_B$ to be at least a factor of $3$ larger
than the ones due to $\overline{H}^{(2)}$ for the maximum values
of $\Gamma_0$ explored.

\begin{figure}[t]
\includegraphics[width=3.1in]{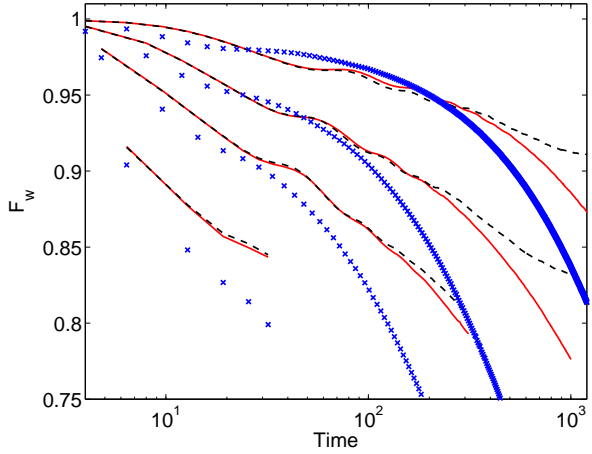}
\caption{(Color online) Bath dynamics effect for PCDD$_2$ with
$\Gamma_0=0$ (black dashed lines), $0.01$ (red solid lines), and
$0.1$ (blue crosses). The parameters are $N=20$ and $\tau=0.2$,
$0.25$, $0.3$, and $0.4$
from top to bottom.} \label{fig:bd}    
\end{figure}

The results are summarized in Fig.~\ref{fig:bd}, where a
two-dimensional $4 \times 5$ QD with nearest neighbor-intrabath
coupling is considered. On one hand, the performance of PCDD$_2$
is affected only slightly for sufficiently small $\tau$ and
$\Gamma_0$, as expected.  Note that for a standard GaAs QD, the
characteristic time scale for the nuclear spin dynamics is $\sim
100\mu$s, implying that $\Gamma_0\ll 0.01$ in our model. Thus,
results obtained for $\Gamma_0=0$ are applicable to typical GaAs
QDs. On the other hand, as it is also clear from
Fig.~\ref{fig:bd}, the long-time PCDD$_2$ fidelity deteriorates
significantly in the presence of a sufficiently fast bath,
especially for larger pulse delay $\tau$. Based on the spin-chain
results of Ref. \onlinecite{Santos06}, randomized protocols such
as SRPD are capable to be, on average, more robust to the effects
of the underlying bath evolution.

\subsection{Magnetic bias field and initial bath polarization}
\label{sec:bp}

\begin{figure}[t]
\includegraphics[width=3in]{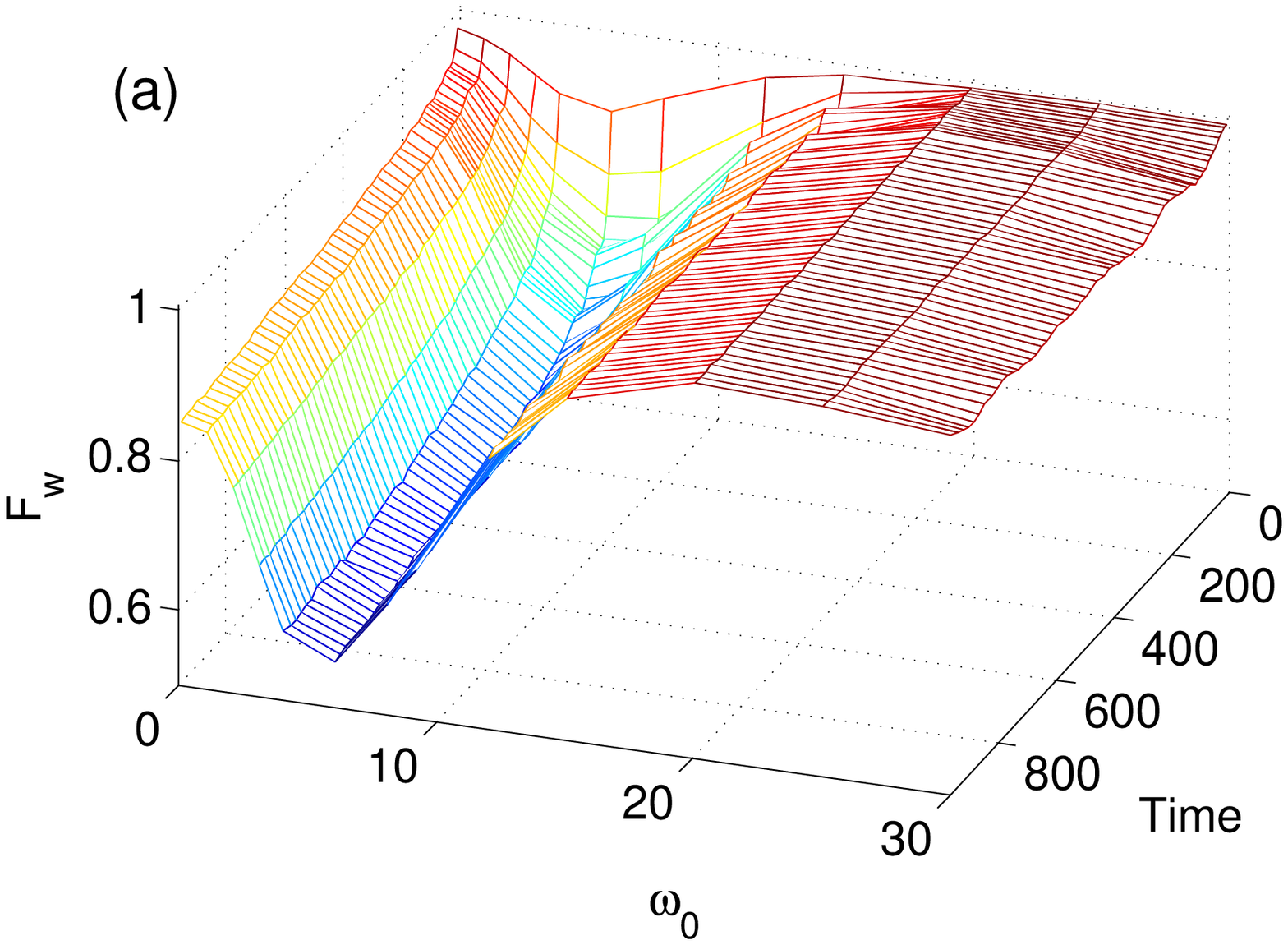}
\includegraphics[width=3in]{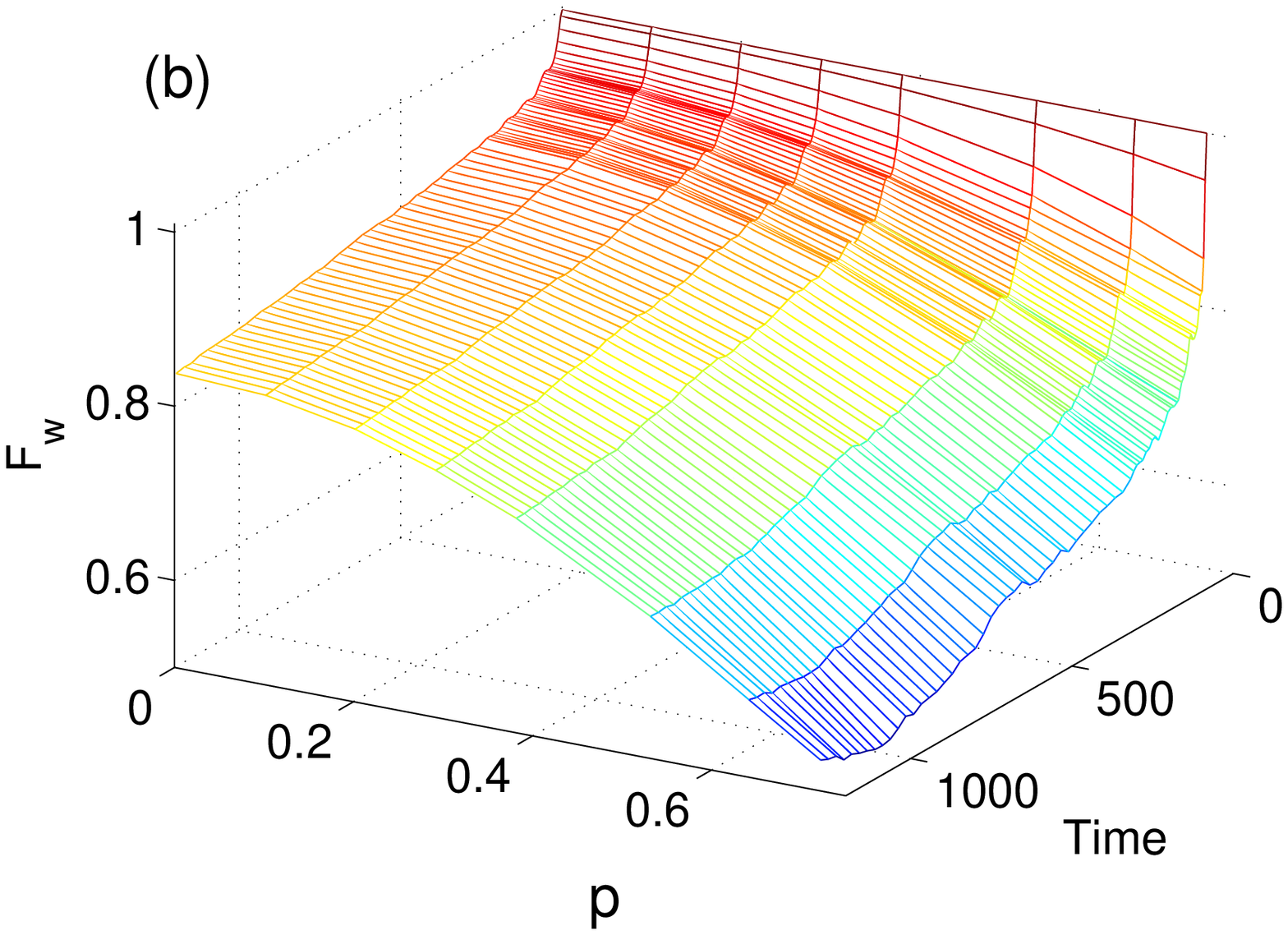}
\caption{(Color online) Effect of bias magnetic field for different
evolution times (top) and effect of initial bath polarization for
(bottom) on the worst-case performance
of PCDD$_2$. Parameters are $\tau=0.3$ and $N=15$. All energies
are measured in units of $A$.}
\label{fig:bmf} 
\end{figure}
In practice, a non-zero Zeeman splitting in Eq.~(\ref{Hel}) might
either be desirable in order to impose a controllable quantization
axis on the electron spin and/or in any case result from residual
magnetic fields in the device. As numerical results show (see
Fig.~\ref{fig:bmf}(a)), the dependence of the PCDD$_2$ worst-case
performance is non-monotonic with $B_0$, by decreasing with small
magnetic fields and improving for large magnetic fields -- this effect
being more prominent for long evolution times. While this behavior is qualitatively consistent with the emergence of a predominantly
pure-dephasing process in the presence of a strong bias field, the
latter also leads, in general, to temporal modulation
of the fidelity, similar to the effect of electron spin echo
envelope modulation~\cite{Rowan65}.

Interestingly, a non-zero initial polarization of the
bath spins acts in a similar manner, see Fig.~\ref{fig:bmf}(b). In
these simulations, we assumed that the initial state of the bath
is described by a density matrix of the form
$$\rho_B(0) =\frac{1}{Z_\beta}e^{-\beta \sum_{k=1}^N I_k^z},
\;\;\; Z_\beta = [2\cosh(\beta/2)]^N , $$
\noindent
where $\beta$ denotes as usual the inverse temperature. Accordingly,
the bath polarization is defined as the ratio
$$p = \frac{1}{(N/2)} \sum_{k=1}^N \langle I_k^z\rangle = -
\tanh(\beta/2), $$ \noindent with $\langle\cdot\rangle$ denoting
the expectation over the above bath spin state.  It has been shown
earlier~\cite{Zhang06} that for the free decoherence dynamics of
an electron spin in a QD, a non-zero Overhauser field produced by
a small nuclear-spin polarization is essentially equivalent to an
external bias field. Here, a similar equivalence emerges for the
worst-case performance of PCDD$_2$: the degradation trend of the
PCDD$_2$ fidelity for small polarizations seen in
Fig.~\ref{fig:bmf}(b) resembles that occurring for small bias
magnetic fields, $\omega_0\lesssim 4$. This indicates that the
effect of the Overhauser field of a weakly polarized bath is also
equivalent to that of an external bias field in the presence of DD
pulses. We are unable to further explore this equivalence for
larger values of the initial polarization, due to the fact that
exact simulations with a highly polarized spin bath require
numerical techniques which are beyond our current
capabilities~\cite{Zhang07r}.  On physical grounds, we expect that
the degradation trend should stop, and improved performance should
emerge as $p$ approaches $1$, since the fidelity should achieve
100$\%$ for a fully polarized initial bath state (even in the
absence of control pulses, for appropriate electron spin
alignment).


\section{Control Resources}
\label{sec:ci}

We conclude our analysis by addressing the main simplifying
assumptions and requirements implicit in the control capabilities
invoked so far, with respect to relevant practical constraints.

\subsection{Effect of pulse imperfections}

In all simulations presented thus far, control pulses have been
assumed to exactly achieve a rotation angle of $\phi=\pi$ in no
time -- corresponding to zero width and infinite strength. In
reality, pulses are clearly finite in both strength and width, and
the implemented rotation angle typically deviates from the
intended value due to both systematic control faults and
stochastic parameter fluctuations. While a fully accurate error
modelling is necessarily dependent upon the details of the
physical implementation, our aim in what follows is to gain a
sense on how stable PCDD$_2$ performance is in the presence of
different representative control errors.

\begin{figure}[t]
\includegraphics[width=3in]{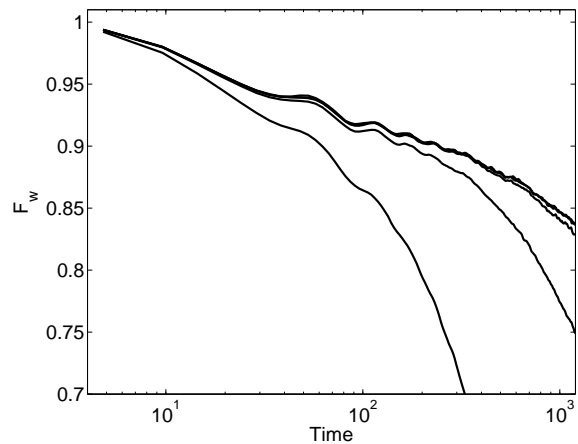}
\caption{Effect of systematic rotation-angle errors on the worst-case
PCDD$_2$ performance, with $\tau=0.3$ and $N=15$.  The relative flip
angle errors are $\varepsilon=0$, $0.001$ (indistinguishable from
$\varepsilon=0$), $0.003$, $0.01$, $0.03$, from top to bottom.}
\label{fig:ae} 
\end{figure}

The effect of a {\em systematic} error in the rotation angle may
be modelled by assuming that $\phi=\pi(1-\varepsilon)$, while
keeping the pulse instantaneous, and letting $\varepsilon \in
[0,1]$ represent the relative flip-angle error.
Figure~\ref{fig:ae} summarizes numerical results for various error
strengths.  The protocol performance is not very sensitive to this
type of error, especially for small $\varepsilon$ (as expected).
In fact, the electron coherence time is still two orders of
magnitude longer than $T_2^*$ even for $\varepsilon=0.03$, which
corresponds to $\sim 5.4^\circ $. With phase compensation
techniques~\cite{Slichter92,Haeberlen76}, DD performance might be
further improved.

\begin{figure}[t]
\includegraphics[width=2.96in]{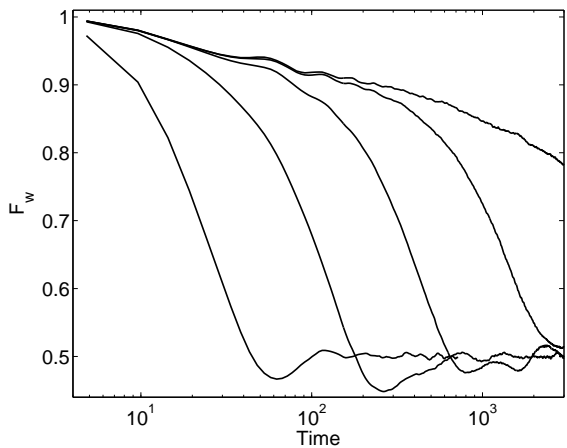}
\caption{Effect of finite pulse width on the worst-case PCDD$_2$
performance, with $T_c=4.8=16\tau+20w$ and $N=15$. The pulse
widths are $w=0$, $0.0024$, $0.008$, $0.024$, $0.08$,
from top to bottom.} \label{fig:pw}    
\end{figure}

The effect of finite pulse durations may be accounted for by
assuming that each pulse is rectangular, with the width $w$ and the corresponding strength $\pi/w$ adjusted to represent an ideal $\pi$
pulse for an isolated spin.  That is, the relevant control Hamiltonian
becomes
$${H'}_c(t) = \sum_\ell \frac{\pi}{w}
({\bm S}\cdot \hat {\bm n}(t)) [h(\ell\tau-t) -
h(\ell\tau-t-w)],$$ \noindent where $h(\cdot)$ is the Heaviside
step function. In simulations, we have considered the width of the
pulse to be up to one third of the pulse delay, $w\leq \tau/3$.
Figure~\ref{fig:pw} summarizes results on the worst-case PCDD$_2$
performance for different values of $w$: Clearly, DD performance
depends quite sensitively on the pulse width. For a typical QD
with decoherence time $T_2^*\sim 10$ ns, this means that pulse
delays $\tau \sim 4$ ns and pulse widths no longer than $w=0.024$
(about $0.3$ ns) are required in order to extend $T_2^*$ by a
factor of $\sim 100$ under PCDD$_2$.

Improved control design is needed to relax such stringent limitations,
in particular to ensure that high-fidelity DD may be achieved with finite
bandwidth.  While further analysis along these lines is beyond our
current scopes, preliminary results indicate how an approach based on
{\em Eulerian control}~\cite{Euler,Viola04a} may allow pulse widths as
long as $\tau$ to be employed \cite{Taylor}.  A yet more sophisticated
approach is to resort to numerical pulse shape optimization, for
instance based on the recently proposed open-system gradient ascent
pulse engineering algorithm~\cite{Grape,Wilhelm}.

\subsection{Feasibility considerations}

From a practical standpoint, the main requirements for DD
implementations are the ability to effect sufficiently fast
single-spin rotations -- along two orthogonal axes in an appropriate
frame if storage of {\em arbitrary} electron spin states is sought.
While this is a highly non-trivial task, single-spin rotations have by
now been experimentally realized in both gate-defined
radiofrequency-controlled
GaAs double QDs \cite{Koppens06} and self-assembled singly-charged
(In,Ga)As/GaAs QDs \cite{Greilich06,Dutt06,Chen04} -- proposals for
further improving rotation speed and gating time being actively
investigated in parallel \cite{Emary07,Coish07}.  Likewise, multipulse
CPMG protocols have been successfully implemented on single impurity
centers in a solid-state matrix
\cite{Fraval05,Fraval04,Childress06,HansonNV06}. In typical ESR
settings, for instance, magnetic pulses as narrow as $20$~ns and
gating times of the order of $100$ ns are currently attainable for
typical GaAs QDs at dilution refrigerator temperatures
\cite{Rimberg06}.

While the above-mentioned accomplishments and figures give hope that
full-fledged DD experiments in QDs should become accessible in a near
future, some additional remarks may be useful in connection with the
prospective relevance of our results to different control
implementations.  Since, as remarked, our main focus has been on the
zero-field limit ($\omega_0=0$), standard ESR techniques are not
directly viable to effect the intended rotations.  Rather, the control
we are envisioning is based on {\em direct magnetic switching}, which
may be accomplished in principle by having access to two independent
inductances oriented along perpendicular axes. In order for the
required control to be achievable via ESR or coherent Raman
spectroscopy in the presence of a permanent static field, pulse widths
and separations in the (sub)ns range would imply a resonance frequency
$\omega_0 \gtrsim 1$ GHz. In these conditions, as mentioned, the
hyperfine decoherence process would be largely dephasing-dominated, and
application of two-axis DD sequences such as PCDD$_2$ would result in
even better performance than at zero field -- similar to the
conclusion reached in Sec.  \ref{sec:bp}.  As also noticed in Ref.
\onlinecite{Bergli06}, a potential advantage of control techniques not
relying on the presence of an external field, however, is applicability
over a wider range of parameters, along with the possibility to avoiding
perturbations on the spin dynamics in between control operations.
From this perspective, a more careful feasibility study of direct
switching schemes in QD devices appears well-worth pursuing.


\section{Conclusion}
\label{sec:conclusion}

We have provided an in-depth quantitative analysis of
the dynamical decoupling problem for a central spin system in the
zero-field limit.  While our main intended application has been
long-time preservation of electron spin coherence in a quantum dot by
suppression of hyperfine-induced coupling, we expect our main methods
and conclusions to be relevant for control of nanospin systems
described by a similar model Hamiltonian. Our main conclusions may be
summarized as follows.

For short evolution times and inter-pulse delays $\tau$ which obey the
condition $\omega_c \tau\lesssim 1$, $\omega_c$ being the spectral
{\em cut-off frequency of the total system}, analytical results based
on average Hamiltonian theory and the Magnus expansion are in
excellent agreement with the results of exact numerical simulations.
For longer evolution times and pulse separations, which are beyond the
domain of applicability of analytical approaches, DD can still ensure
high-fidelity preservation of arbitrary electron spin states, as long
as typical control time scales are short with respect to the {\em
spectral width of the total system}.

If knowledge of the initial electron state is available, cyclic DD
protocols capable to completely freeze electron decoherence in the
long-time limit may be designed -- the attainable coherence value
depending on both the protocol and the pulse delay.

By studying the effect of important experimental factors and control
non-idealities -- including the effect of intrabath interactions,
external magnetic fields, and typical systematic pulse errors --
we conclude that even imperfectly implemented decoupling protocols
may still be able to significantly prolong the electron coherence
time.

\acknowledgments

It is a pleasure to thank David G. Cory and Alexander J. Rimberg for
useful discussions. Work at the Ames Laboratory was supported
by the Department of Energy --- Basic Energy Sciences under
Contract No. DE-AC02-07CH11358, and by the NSA and ARDA under Army
Research Office contract DAAD 19-03-1-0132. Part of the calculations
used resources of the National Energy Research Scientific Computing
Center, which is supported by the Office of Science of the
U. S. Department of Energy under Contract No. DE-AC02-05CH11231. L. V.
also gratefully acknowledges partial support from the NSF through
Grant No. PHY-0555417, as well as hospitality and partial support from
the {\em Center for Extreme Quantum Information Theory} at MIT during
the final stages of this work.

\appendix

\section{Analytical Study of Decoherence Freezing Under CPMG Decoupling}
\label{sec:cpmg}

The simplicity of the CPMG ($\tau Z\tau Z$) protocol allows for an
analytical study of the saturation effect discussed in
Sec.~\ref{sec:fb}. This will be done below via the QSA~\cite{Taylor06,
Zhang06}, as well as based on a semiclassical random field model.

\subsection{Quasi-Static Approximation}

Within the QSA~\cite{Taylor06, Zhang06}, the Hamiltonian
Eq.~(\ref{eq:h}) reduces to $H_A$ Eq.~(\ref{eq:ha}). Let ${\mathbf
I}=\sum_k {\mathbf I}_k$, and $M=I_z$. Due to the symmetry of
$H_A$ and the CPMG protocol, ${\mathbf I}^2$ and $M+S_z$ are
constants of motion. Let $|\Psi(0)\rangle= |\uparrow\rangle
\otimes |I,M\rangle$ denote the the joint initial state, where the
first ket-vector corresponds to the state of the electron spin and
the second ket-vector denotes the state of the bath. Then the
evolution of the controlled system only couples the pairs of
levels $|\uparrow\rangle \otimes |I,M\rangle$ and
$|\downarrow\rangle \otimes |I,M+1\rangle$.  (Similar
considerations apply to the case where the electron spin is down,
initially).

After $n$ CPMG cycles, the evolution operator is
\begin{eqnarray*}
U(2n\tau) &=& \left(\begin{array}{cc}d_1
    &d_2^*\\d_2&-d_1^* \end{array}\right)^{2n},
\end{eqnarray*}
where we have defined
\begin{eqnarray*}
d_1 &=& \cos{\Omega \frac{\tau}{2}} - i{\frac{B}{\Omega}}
    \sin{\Omega \frac{\tau}{2}}, \\
d_2 &=& -i{\frac{C}{\Omega}} \sin{\Omega\frac{\tau}{2}},
\end{eqnarray*}
with $C=A\sqrt{(I-M)(I+M+1)}$, $B=A(M+1/2)$, and $\Omega^2 =
B^2+C^2$.  In the best-case scenario, the overall system is
initialized in the state
$$ |\Psi(0)\rangle_b = |\uparrow\rangle \otimes |I,M\rangle,$$
\noindent
whose survival probability is given by
\begin{eqnarray*}
|\langle \Psi(0)|\Psi(2n\tau)\rangle|^2 &=& 1- {\frac{C^2}{B^2}}
\tan^2\theta \cos^2 2n\theta,
\end{eqnarray*}
where $\tan\theta=d/\sqrt{1-d^2}$ with $d = {\text{Im}}(d_1)$.

The initial bath state is a completely mixed state and can be
rewritten in the basis of $|I,M\rangle$ as
\begin{eqnarray*}
\rho_B(0) &=& \sum_{I,M} P(I,M) |I,M\rangle\langle I,M|,
\end{eqnarray*}
with $P(I,M)\simeq (I /D\sqrt{2\pi D}\;)e^{-I^2/2D}$ for large $N$ and
$D=N/4$~\cite{Melikidze04}. Averaging over the nuclear spin states, we
obtain the survival probability, the input-output fidelity, at time
$T=2n\tau$,
\begin{eqnarray*}
F(2n\tau) &\hspace*{-1mm}=\hspace*{-1mm}& 1- \int dI dM P(I,M) {\frac{C^2}{B^2}} \tan^2\theta \cos^2 2n\theta \\
&\hspace*{-1mm}=\hspace*{-1mm}
& 1- \frac{1}{2}\int dI dM P(I,M) {\frac{C^2}{B^2}} \tan^2\theta ( 1
+ \cos 4n\theta) \nonumber \\
\end{eqnarray*}

For $n$ large enough so that $4n\theta\gg 1$, we may safely
neglect the contribution from rapidly oscillating part $\cos
4n\theta$.  In these conditions, $F(2n\tau)$ becomes
time-independent and decoherence is frozen,
\begin{eqnarray*}
F(2n\tau) \rightarrow F_f &=& 1 - \frac{1}{2}\int dI dM P(I,M)
{\frac{C^2}{B^2}} \tan^2\theta. \nonumber \\
\end{eqnarray*}
In the limit of small $\tau$, this yields
$$F_f = 1- \frac{1}{16} \tau^2 A^2 N = 1-\frac{\tau^2}{2T_2^{*2}}.$$
\noindent
For randomly distributed $A_k$, $A^2N \mapsto \sum_k A_k^2$.

For a single control cycle ($n=1$) as considered in Sec. \ref{fb1},
and sufficiently small $\tau$ so that $\theta \ll 1$, it is
straightforward to find that
\begin{eqnarray*}
F(T_c) &=& \frac{4}{5} -
\frac{1}{15}\Big[(1-4D\tau^2)e^{-2D\tau^2}  \nonumber \\
& -& 4(1-D\tau^2)e^{-D\tau^2/2} \Big] \nonumber \\
&\approx & 1-{\frac{D^2}{2}} \tau^4 + {\mathcal O}(\tau^6)
= 1- \kappa \tau^2 T_c^2 + {\mathcal O}(\tau^4 T_c^2),
\end{eqnarray*}
consistent with the result reported in the main text.

\subsection{Classical Random Field Model}

Yet another simple way to describe the saturation associated with the
CPMG protocol at long times is as follows.  In situations where the
back-action effects from the system into the bath may be neglected, a
plausible approximation is to treat the environment as a classical
random field.  This translates into rewriting the coupling Hamiltonian
as
$$ H=\vec{B}\cdot \vec{\sigma}\;, $$
where the effects of the randomly fluctuating field $\vec{B}$,
\begin{eqnarray}
&& B_x = B \cos \theta \sin \phi, \nonumber \\
&& B_y = B \sin \theta \sin \phi, \nonumber \\
&& B_z = B \cos \phi, \nonumber
\end{eqnarray}
are taken into account by averaging over the entire Bloch sphere
of the system spin, $0\leq \phi \leq \pi$ and $0\leq \theta \leq
2\pi$.

Under CPMG,
the unitary evolution after the completion of $n$ cycles may be
written as
\begin{eqnarray*}
U(2n\tau) &\hspace*{-1mm} = \hspace*{-1mm}
&\left( e^{-i (-B_x X - B_y Y + B_z Z) \tau }
e^{-i (B_x X + B_y Y + B_z Z) \tau} \right)^n \nonumber \\
&\hspace*{-1mm}=\hspace*{-1mm}&V\left( \begin{tabular}{cc}
$e^{-i\lambda n}$ & 0 \\
0 & $e^{i\lambda n}$
\end{tabular} \right)V^{-1} \; ,
\end{eqnarray*}
where $V$ is the matrix of eigenvectors of $U(T_c)$ and $\exp(\pm i
\lambda )$, $\lambda \in {\mathbb R}$ denote the corresponding
eigenvalues.  For a particular initial state aligned with the dominant
term of the AHT, that is $\overline{\cal H}^{(0)}=B_z Z$, the fidelity
after $n$ cycles becomes
$$ F_b(2n\tau) = \frac{\cos(B \tau)^2 + \sin (B \tau)^2 \sin(\phi)^2
(1 - \sin(n\lambda )^2)} {\cos(B \tau)^2 + \sin (B \tau)^2
\sin(\phi)^2} \;.$$
\noindent
In the long time limit, $n=t/(2\tau) \rightarrow \infty$, and upon
averaging over $\theta$ and $\phi$, we finally obtain
$$ F_b \rightarrow F_f = 1 - \frac{B^2 \tau^2}{3} \;, $$
in agreement with the expression quoted in the main text.

Notice that the same result may also be obtained by restricting the
analysis to the two dominant terms in the AHT (that is,
$\overline{\cal H}^{(0)}, \overline{\cal H}^{(1)}$)
and by writing the propagator after $n$ cycles as
$$ U(t=2n\tau) = e^{-i[B_z Z t + B_z (B_x Y - B_y X) t \tau ] }. $$
\noindent
While in principle no {\em a priori} reason exists to expect such
AHT-based description to yield the correct answer, a similar treatment
successfully describes long-time magnetization ``pedestals" in NMR
decoupling \cite{Haeberlen76}.



\begin{thebibliography}{78}
\bibitem{Xremark} The fact that the $x$ axis in preferred in the
induced bath Hamiltonian (\ref{renHB}) reflect the structure of
the base sequence $C_1$ chosen for concatenation, which causes $X$
rotations to be comparatively more frequent, see Eq. (\ref{cddcycle}).
\bibitem{Nremark} Since the order of the leading decoherence terms is
independent of $N$, we use the smallest possible bath size, $N=2$, in
implementing the expansion.
\end{thebibliography}
\end{document}